\documentclass[conference]{IEEEtran}
\IEEEoverridecommandlockouts
\usepackage{cite}
\usepackage{amsmath,amssymb,amsfonts}
\usepackage{graphicx}
\usepackage{textcomp}
\usepackage{multicol}
\usepackage{multirow}
\usepackage{hyperref}
\usepackage{algorithm}
\usepackage{algpseudocode}
\usepackage{graphics}
\usepackage{booktabs}
\usepackage{verbatim}
\usepackage{fancyvrb}
\usepackage{fvextra} 
\usepackage{enumitem} 
\usepackage{xcolor}
\def\BibTeX{{\rm B\kern-.05em{\sc i\kern-.025em b}\kern-.08em
    T\kern-.1667em\lower.7ex\hbox{E}\kern-.125emX}}
    
\begin{document}

\title{Customized Retrieval-Augmented Generation with LLM for Debiasing Recommendation Unlearning\\
}
\author{
    \IEEEauthorblockN{Haichao Zhang,
                      Chong Zhang,
                      Peiyu Hu,
                      Shi Qiu,
                      Jia Wang\textsuperscript{\rm $^{\dagger}$} \thanks{$^{\dagger}$ Corresponding Author.}}
    \IEEEauthorblockA{Xi'an Jiaotong-Liverpool University\\
                      \textbf{Email:} haichao.zhang22@student.xjtlu.edu.cn, jia.wang02@xjtlu.edu.cn}
}

\maketitle

\begin{abstract}

Modern recommender systems face a critical challenge in complying with privacy regulations like the “right to be forgotten”: removing a user’s data without disrupting recommendations for others. Traditional unlearning methods address this by partial model updates, but introduce propagation bias—where unlearning one user’s data distorts recommendations for behaviorally similar users, degrading system accuracy. While retraining eliminates bias, it is computationally prohibitive for large-scale systems. 
To address this challenge, we propose CRAGRU, a novel framework leveraging Retrieval-Augmented Generation (RAG) for efficient, user-specific unlearning that mitigates bias while preserving recommendation quality. CRAGRU decouples unlearning into distinct retrieval and generation stages. In retrieval, we employ three tailored strategies designed to precisely isolate the target user’s data influence, minimizing collateral impact on unrelated users and enhancing unlearning efficiency. Subsequently, the generation stage utilizes an LLM, augmented with user profiles integrated into prompts, to reconstruct accurate and personalized recommendations without needing to retrain the entire base model.
Experiments on three public datasets demonstrate that CRAGRU effectively unlearns targeted user data, significantly mitigating unlearning bias by preventing adverse impacts on non-target users, while maintaining recommendation performance comparable to fully trained original models. Our work highlights the promise of RAG-based architectures for building robust and privacy-preserving recommender systems. 
The source code is available at: \href{https://github.com/zhanghaichao520/LLM_rec_unlearning}{https://github.com/zhanghaichao520/LLM\_rec\_unlearning}.

\end{abstract}


\section{Introduction}



Recommender systems (RS) rely heavily on user-generated data to deliver personalized experiences~\cite{he2020lightgcn, liu2023federated, hu2008collaborative}, raising concerns over privacy and data integrity. 
Users now demand the “right to be forgotten” under regulations like GDPR~\cite{gdpr2016general}, while poisoned or outdated data further threaten model quality~\cite{huang2021data}. 
These challenges have driven growing interest in \textit{recommendation unlearning}—removing specific user influences from trained models while preserving overall utility and efficiency.

\begin{figure}[ht]
  \centering
  \includegraphics[width=.5\textwidth]{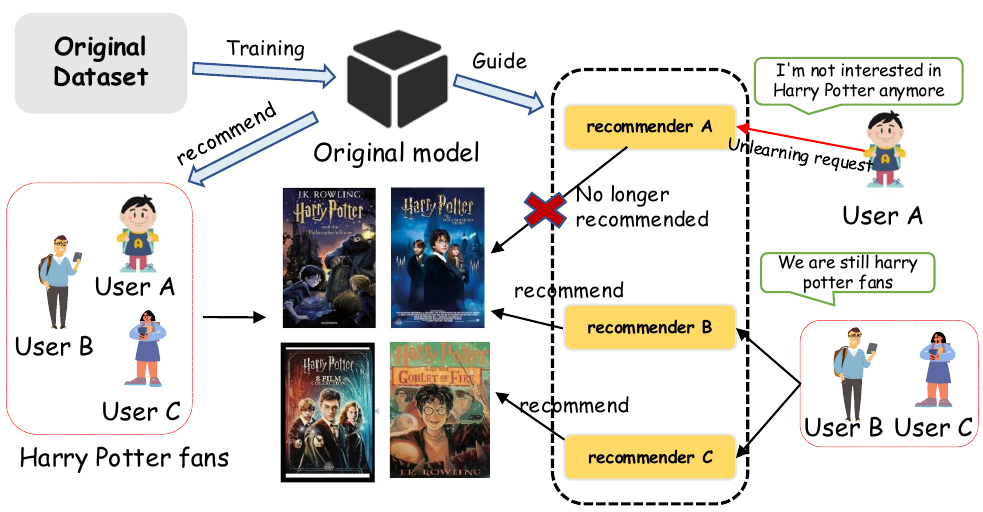}
  \vspace{-10pt}
  \caption{Traditional methods use a single shared recommendation model for all users, where unlearning one user's data alters global parameters, potentially degrading recommendations for others. In contrast, our method leverages Retrieval-Augmented Generation (RAG) with LLMs to perform efficient and precise user-level unlearning without affecting unrelated users.}
  \vspace{-10pt}
  \label{fig:intro}
\end{figure}
The most straightforward unlearning approach, retraining the model from scratch on the remain dataset, guarantees complete removal but incurs prohibitive computational costs, especially for large-scale system~\cite{wang2024comprehensive, chen2021machine}. To mitigate this, two primary classes of methods have been developed. Exact unlearning methods, often based on a partition-and-retrain framework like SISA~\cite{sachdeva2024machine}, aim to isolate changes by retraining only affected sub-models, with extensions like GraphEraser~\cite{chen2022graph} and RecEraser~\cite{chen2022recommendation} for recommendation scenarios.  Alternatively, approximate unlearning methods aim for efficiency by estimating and reversing the impact of data to be forgotten. Among these, IFRU~\cite{zhang2024recommendation,chen2024post}, and SCIF~\cite{li2023selective} have been proposed
to approximate the impact of individual training data points via influence function. 
However, these methods often struggle with the computational burden of calculating Hessian matrices for large models~\cite{koh2017understanding}. Despite the distinct mechanisms of these exact and approximate approaches, a pervasive challenge known as \textbf{unlearning bias} can emerge, undermining their practical utility.


This \textbf{unlearning bias} refers to the unintended and detrimental impact on recommendation quality for remain users by the unlearning request. It arises because the influence of the user being “forgotten” is often entangled with remaining users due to collaborative effects~\cite{li2024survey}. However, both primary paradigms of unlearning can suffer from this:
Exact unlearning methods (e.g., sharding) introduce bias as data removal alters sub-models, negatively impacting co-located users in the same shard~\cite{li2025causal,sachdeva2024machine}. Similarly, approximate unlearning methods, such as influence functions, may inadvertently shift the embeddings of behaviorally similar users, leading to degraded recommendations for these remaining users. As illustrated in \autoref{fig:intro}, if User A, a Harry Potter fan, requests their data be unlearned, these methods might degrade recommendations for other Harry Potter fans (Users B and C). Beyond quality, existing partition-aggregation unlearning methods face efficiency limitations. Liu et al.~\cite{liu2022forgetting} show small unlearning requests often necessitate retraining nearly all sub-models, rendering these methods computationally impractical under continuous or high-volume unlearning scenarios.

To overcome these limitations, we propose CRAGRU, a novel approach that reframes unlearning as a targeted information retrieval problem within a Retrieval-Augmented Generation (RAG) architecture, leveraging Large Language Models (LLMs). This RAG paradigm is crucial because it allows unlearning to be primarily implemented at the retrieval stage, where specific information fed to the LLMs can be precisely filtered. This performs user-level atomic operations, isolating unlearning effects to the targeted user. CRAGRU operates through three tightly coupled stages: First, during retrieval, user interactions, profiles, and item metadata are structured as natural language prompts. This stage is paramount for unlearning, as it systematically filters out data that should be forgotten from the retrieval index, ensuring only relevant, non-forgotten information influences subsequent steps. To optimize this process for unlearning efficacy and efficiency, we introduce three novel strategies: (1) User preference-based retrieval, which selects the most representative interactions aligned with long-term user preferences (e.g., sustained genre interests); (2) Diversity-aware retrieval, balancing item diversity with recommendation performance; and (3) Attention-aware retrieval, which identifies high-impact user interactions based on transformer attention weights. Next, in augmentation, candidate items from traditional backbone models (e.g., LightGCN~\cite{he2020lightgcn}) are refined by the LLMs, fusing collaborative signals with semantic knowledge—for instance, inferring that “users who like Harry Potter also enjoy fantasy world-building” to enhance candidate relevance. Finally, during generation, the LLM synthesizes personalized recommendations solely from these augmented and meticulously filtered candidates, ensuring that the unlearning process has minimal negative impact on other related users, thereby achieving precise, debiased recommendation unlearning. 
We summarize the main contributions of this paper as follows:
\begin{itemize}[itemsep=0.5pt,parsep=0pt,partopsep=0pt,topsep=1.0pt]
    \item To the best of our knowledge, CRAGRU is the first framework to unify retrieval-augmented LLMs with traditional recommenders for unlearning. By treating each user’s recommendations as an atomic unit, it achieves minimal impact on non-target users and efficient unlearning without retraining or parameter updates.
    \item We design three novel interaction retrieval mechanisms to balance unlearning efficacy and recommendation quality. Specifically, Preference-aware retrieval (the most representative interactions); Diversity-constrained retrieval (item coverage); and  Attention-guided retrieval (identify the important user interactions ).
    \item Extensive validation across three datasets and two backbone models (LightGCN, BPR). CRAGRU reduces the average unlearning time by 4.5× versus SOTA baselines, while retaining approximately 90\% of the recommendation model’s performance before unlearning.
\end{itemize}

\section{RELATED WORKS}

\subsection{Machine Unlearning}
Machine unlearning is designed to remove the impact of a specific subset of training data from a trained model~\cite{nguyen2022survey}. 
A direct approach is to update the dataset and retrain; however, this will incur a significant computational overhead. Initial research focused on traditional machine learning tasks~\cite{tarun2023deep,brophy2021machine}.
For example, efficient data deletion of $K$-means clustering~\cite{ginart2019making}, incremental and decremental learning algorithms for linear support vector machines~\cite{cauwenberghs2000incremental,karasuyama2010multiple}, and fast Bayes data deletion based on statistical query learning~\cite{cao2015towards}.
However, due to the limited application scenarios and lack of generalization of these methods, it is difficult to apply them to non-convex models such as deep neural networks with huge parameter spaces.
To improve the generalization of unlearning, Bourtoule et al. proposed SISA~\cite{bourtoule2021machine}, a model-agnostic unlearning framework. The core idea is to divide the original dataset into multiple shards, train sub-models with these shards, and finally summarize the sub-model results. During the unlearning process, only the sub-models of the labeled shards need to be retrained.
Subsequently, chen et al. proposed GraphEraser~\cite{chen2022graph}, which enhanced the data partitioning method for graph-based datasets based on SISA and improved the performance of model unlearning in graph data scenarios.
As large language models continue to develop, some studies have begun to utilize these models for unlearning~\cite{chen2023unlearn,yao2023large}. For instance, Chen et al. introduced lightweight unlearning layers within transformers, utilizing a selective teacher-student objective. Their method has demonstrated effective performance in classification and generation tasks~\cite{chen2023unlearn}.

\subsection{Recommendation Unlearning}
Recommendation Unlearning aims to enable the model to ``forget" information about specific users or items to satisfy privacy protection requirements or facilitate model updates.
Existing methods can be broadly categorized into the following types: RecEraser~\cite{chen2022recommendation} retains collaborative information through data partitioning and aggregation; however, adhering to the SISA~\cite{bourtoule2021machine} paradigm restricts both performance and efficiency.
Unlearn-ALS~\cite{xu2023netflix} introduces a fine-tuning optimization approach tailored for bilinear models. AltEraser\cite{liu2022forgetting} breaks down the Unlearning problem into multiple sub-problems to simplify computation. FRU~\cite{yuan2023federated} concentrates on unlearning within federated recommendation systems, striving to eliminate the influence of particular users. Additionally, some studies employ influence functions to estimate the impact of data on the model, enabling rapid updates without the need for retraining, as seen in IFRU~\cite{zhang2024recommendation}. Furthermore, research has explored unlearning in various models, including session-based~\cite{xin2024effectiveness} and sequential-based~\cite{ye2023sequence} approaches.


\subsection{LLMs for Recommendation}
Large Language Models (LLMs) have demonstrated immense application potential in recommendation systems, attracting extensive research interest~\cite{zhao2023recommender, lin2023can, liu2023pre, wu2024survey, shu2024knowledge}. 
Some studies explore using LLMs as inference models by designing prompts to guide them in performing recommendation tasks. 
For example, P5~\cite{geng2022recommendation} leverages item indices to convert user interactions into text prompts for model training, 
while M6-Rec~\cite{cui2022m6} textualizes user behavior data and transforms recommendation tasks into language tasks. M6-Rec achieves efficient recommendation models under limited hardware resources by employing techniques such as enhanced prompt tuning, post-processing interactions, and early exiting. 
Chat-REC~\cite{gao2023chat} translates user profiles and interaction information into prompts to construct conversational recommendation systems, 
whereas InstructRec~\cite{zhang2023recommendation} and TALLRec~\cite{bao2023tallrec} utilize instruction fine-tuning methods to enable LLMs to execute recommendation tasks more effectively. 
Additionally, some research attempts to model the structured relationships between user behaviors and items using LLMs to improve recommendation performance. For instance, LLMRec~\cite{wei2024llmrec} strengthens recommendation systems by adopting three simple yet effective LLM-based graph augmentation strategies. However, directly applying LLMs for recommendations often encounters challenges such as high computational costs and slow inference speeds. 
To address these issues, some studies have adopted alternative approaches, such as combining LLM-based data augmentation methods with classical Collaborative Filtering (CF)~\cite{ren2024representation}, aiming to enhance recommendation performance while ensuring the reliability of the results.

\section{PRELIMINARIES}
\subsection{Problem Formulation}
\label{Problem Formulation}
\subsubsection*{Recommendation Task} In a typical recommendation scenario, we are given a set of users $\mathcal{U} = \{u_1, u_2, ..., u_m\}$ and a set of items $\mathcal{I} = \{i_1, i_2, ..., i_n\}$, where $m$ and $n$ denote the number of users and items, respectively.
The interactions between users and items can be represented as a rating matrix $\mathbf{R} \in \mathbb{R}^{m \times n}$, where $R_{ui}$ denotes the rating given by user $u$ to item $i$. In this work, we consider explicit feedback in the form of ratings ranging from 1 to 5. 
We define the dataset $\mathcal{D}=\{(u,i,r)|u \in \mathcal{U}, i \in \mathcal{I}, r \in \{1,2,3,4,5\}\}$, where $(u,i,r)$ mean that user $u$’s rating for item $i$ is $r$. 
The goal of a recommendation model is to predict the missing ratings in $\mathbf{R}$.
Specifically, given user $u$’s historical interaction records
 $R_u = \{ R_{ui} | i \in \mathcal{I}\}$, the recommendation model aims to learn a prediction function $\hat{R_{ui}} = f(u,i|R_u)$ that can accurately predict the rating of user $u$ for item $i$.
\subsubsection*{Recommendation Unlearning} 
In recommendation systems, the objective of recommendation unlearning is to respond to users' withdrawal requests by eliminating the impact of specific data from the trained model.
Assume user $u$ can submit any data withdrawal request $\mathcal{D}_f^u \subseteq \mathcal{D}_u$ to remove their personal information, where $\mathcal{D}_u$ represents all interactions of user $u$. 
We define the set of data withdrawal requests from users requiring unlearning as $D_f=\bigcup_{u \in U} D_f^u$, representing the dataset that needs to be "forgotten". Correspondingly, the remaining dataset is defined as $D_r = D \setminus D_f$.
The task of recommendation unlearning to obtain an unlearned model $\hat{f}$ trained on $D_r$. As introduced in~\cite{li2024making,li2023ultrare}, this task has the following general goals:
\begin{enumerate}
    \item \textbf{G1: Unlearning completeness.} The model must completely forget the deleted data, ensuring that the removal does not adversely affect the model parameters.

    \item \textbf{G2: Unlearning efficiency.} Given the high computational cost of large-scale recommender models, unlearning methods must ensure time-efficient execution.
    
    \item \textbf{G3: Model Utility.} The unlearned model can achieve comparable recommendation performance to models retrained from scratch.

\end{enumerate}

\subsection{Recommendation Model}

Our approach leverages a traditional recommendation model (e.g., matrix factorization or deep learning-based models) as the backbone to generate initial coarse-ranking scores, followed by the CRAGRU framework for unlearning and re-ranking. By decoupling the forgetting mechanism from the backbone architecture, CRAGRU enables precise removal of targeted user data (e.g., preferences or interactions) without modifying the original model parameters. CRAGRU  ensures unlearning avoids costly backbone retraining while ensuring compatibility with arbitrary recommendation systems.


The backbone model learns a function $f_{\text{b}}: \mathcal{U} \times \mathcal{I} \rightarrow \mathbb{R}$ that predicts the preference score $\hat{r}_{u,i}$ for user $u$ and item $i$:
$
\hat{r}_{u,i} = f_{\text{b}}(u, i; \theta_{\text{b}})
$, where $\theta_{\text{b}}$ represents the model parameters.
For each user $u \in \mathcal{U}$, the backbone model generates predicted scores for all items in $\mathcal{I}$ and selects the Top-$K$ items (e.g. $K=50$) with the highest scores as the candidate item list $\mathcal{I}_u^{\text{cand}}$ for subsequent recommendation:
\begin{equation}
    \mathcal{I}_u^{\text{cand}} = \operatorname{topK}_{i \in \mathcal{I}} (\hat{r}_{u,i}, K).
\end{equation}
Here, $\operatorname{topK}_{i \in \mathcal{I}} (\hat{r}_{u,i}, K)$ denotes the selection of the set of $K$ elements with the highest predicted scores $\hat{r}_{u,i}$ for the user $u$. This candidate list $\mathcal{I}_u^{\text{cand}}$ will be used in subsequent stages by the LLM to generate the final recommendations.

\section{Proposed Framework: CRAGRU}
\begin{figure}[ht]
  \centering
  \includegraphics[width=0.48\textwidth]{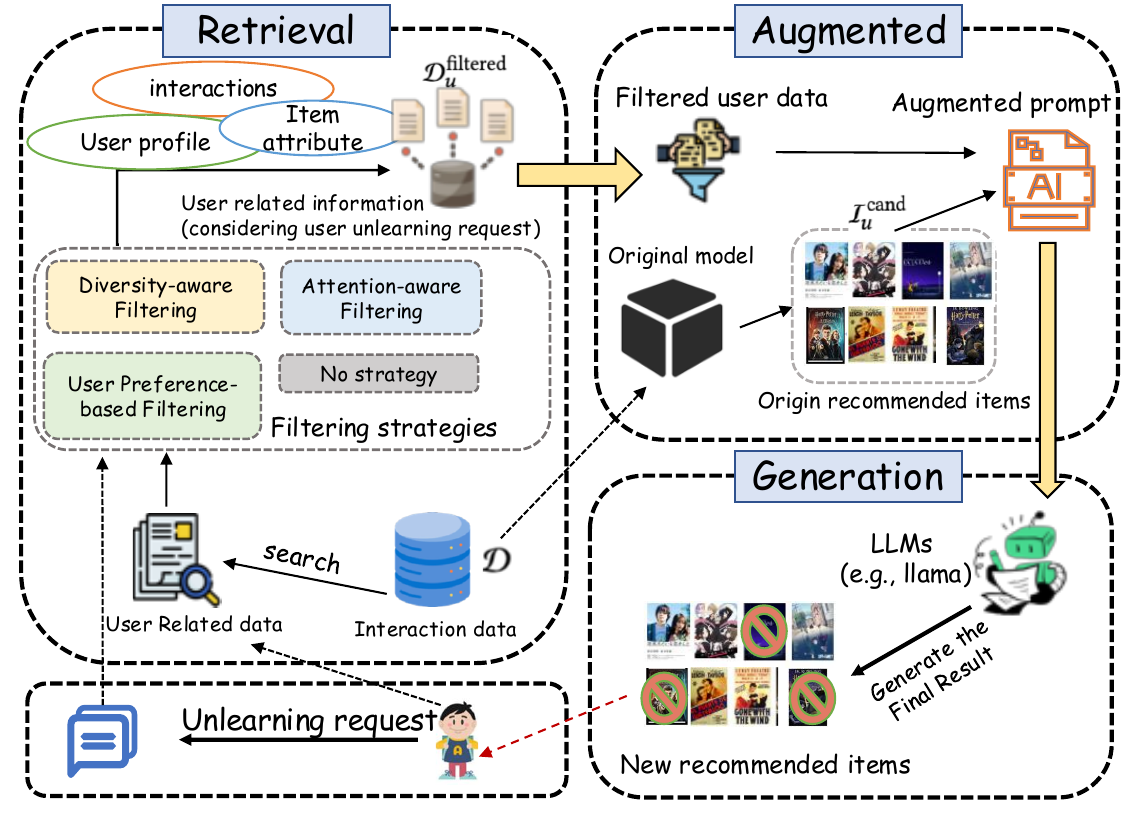}
  \caption{The framework of CRAGRU. When the user submits an unlearning request, the retriever uses the request to fetch relevant information from the dataset and filters out the information that needs to be forgotten. The remaining information is then used to create an unlearning prompt. Finally, the LLM generates recommendation results based on the user's needs.}
  \label{framework}
\end{figure}

The overall framework of CRAGRU is illustrated in \autoref{framework}. 
We propose a RAG-based approach for recommendation unlearning with three stages: \textit{Retrieval, Augmentation,} and \textit{Generation}. 
In the retrieval stage, we design three filtering strategies to flexibly select the key information that influences the user's final recommendation results. By filtering out the user data that needs to be forgotten, we can achieve user-level unlearning. In the Augmentation stage, we combine the key information obtained from the previous stage with the candidate items and auxiliary information to construct the prompt. We then use the prompt generated in the augmentation stage to call the LLM and obtain the recommendation results, thus achieving user-level unlearning.

\subsection{Retrieval with Filtering for Unlearning}
\label{Retrieval}
In this phase, we retrieve user-related information from a dynamically updateable dataset to control the recommendation. 
It is crucial that we filter out the information the user needs to forget and select useful information for recommendations, thereby achieving unlearning at the user level.

As defined before, $\mathcal{D}$ represents the user's historical interaction data. For a given user $u$, we retrieve relevant information: $\mathcal{D}_u = \operatorname{Retrieve}(u; \mathcal{D})$.
Our method enables efficient and accurate user-level unlearning by simply excluding the target data from the retrieval process, thus eliminating the need for retraining.
Therefore, when unlearning is required for certain data $\mathcal{D}_f^{\text{u}} \subseteq \mathcal{D}_u$, we use a filter function to exclude this information from user data, get the filtered dataset: $\mathcal{D}_u^{\text{filtered}} = \operatorname{Filter}(\mathcal{D}_u, \;
\mathcal{D}_f^{\text{u}})$. 
We provide recommendations based on $\mathcal{D}_u^{\text{filtered}}$ to facilitate efficient unlearning. Specifically, we propose three filter strategies to minimize the performance impact of unlearning and enhance unlearning efficiency.

\subsubsection{User Preference-based Filtering}

This strategy aims to select the most representative interaction records from a user's historical interactions based on their inherent preferences, minimizing performance loss during unlearning. The core idea is to categorize items, analyze the proportion of a user's interactions within each category, and sample interactions proportionally.

Specifically, we define the set of categories as $\mathcal{C} = \{c_1,c_2,...,c_n\}$, where n is the total number of categories. There exists a mapping function $f_c: \mathcal{I} \rightarrow \mathcal{C}$, map each item $i \in \mathcal{I}$ to a category $c \in \mathcal{C}$. 
For user $u$, the historical interaction data is $\mathcal{D}_u = \{i_1, i_2, ..., i_m\}$, where $i_j \in \mathcal{I}$ represents the interacted item, and $m$ is the total number of historical interactions for user $u$.
We define $\mathcal{D}_{u,c}$ as the subset of user $u$'s interactions in category $c$:
\begin{equation}
\mathcal{D}_{u,c} = \{i \in \mathcal{D}_u \;|\; f_c(i) = c\}.
\end{equation}

The interaction proportion $p_{u,c}$ of user $u$ in category $c$ is calculated as: $p_{u,c} = \frac{|\mathcal{D}_{u,c}|}{|\mathcal{D}_u|}$, where $|\cdot|$ denotes the cardinality of a set.
Given a target number of interactions to retain $K$ (e.g., $K=100$), the number of interactions $K_c$ to retain for each category $c$ is: $K_c = \lfloor p_{u,c} \times K \rfloor$, where $\lfloor \cdot \rfloor$ is the floor function. 

Finally, we randomly sample $K_c$ interactions from each subset $\mathcal{D}_{u,c}$ to form the filtered interaction set $\mathcal{D}_u^{\text{filtered}}$:
\begin{equation}
\mathcal{D}_u^{\text{filtered}} = \{ x \mid x \in \operatorname{Sample}(\mathcal{D}_{u,c}, K_c), \, c \in \mathcal{C} \},
\end{equation}
where $\operatorname{Sample}(\mathcal{S}, k)$ is a function that randomly samples $k$ elements from set $\mathcal{S}$. 
This preference-based filtering strategy effectively retains the most relevant historical interactions, minimizing the impact on recommendation performance during unlearning. It considers both the overall interaction behavior and the preference distribution across different categories.

\subsubsection{Diversity-aware Filtering}
This strategy addresses the issue where simple filtering based on category proportions can result in interaction data $\mathcal{D}_u^{\text{filtered}}$ being overly concentrated in a few categories. 
We frame it as a resource allocation problem—how to distribute a limited number of interaction records across different categories, balancing item diversity while maximizing overall recommendation performance.

We pre-compute a performance matrix $\mathbf{M}$, where $M[c][p]$ represents the Hit Rate achieved by retaining $p\%$ of interactions from category $c$. Here, $p$ takes values from a predefined set, e.g., $\{10\%, 20\%, ...\}$.

We model this as a knapsack problem, aiming to find the optimal allocation of retention ratios across categories, maximizing overall performance under a fixed total retention ratio (e.g., retaining a total of $K$ interactions, corresponding to a total retention ratio $K'$).
Formally, let $\mathbf{x} = [x_1, x_2, ..., x_n]$ be a vector where $x_c$ denotes the retention ratio for category $c$. Our objective is to find the optimal $\mathbf{x}^{*}$ such that:
\begin{equation}
\mathbf{x}^{*} = \arg\max_{\mathbf{x}} \sum_{c \in \mathcal{C}} M[c][x_c] \quad  \text{s.t.} \quad \sum_{c \in \mathcal{C}} x_c = K'. 
\end{equation}

We employ dynamic programming to solve this optimization problem. Define $DP[i][j]$ as the maximum sum of Hit Rates considering the first $i$ categories with a total retention ratio of $j$. The state transition equation is:
\begin{equation}
DP[i][j] = \max_{x_i \in \{10\%, 20\%, ...\} \cap [0, j]} \{DP[i-1][j - x_i] + M[i][x_i]\},
\end{equation}
where $DP[0][j] = 0$ for all $j$. The final result, $DP[n][K']$, represents the maximized total Hit Rate, and backtracking yields the optimal allocation $\mathbf{x}^{*}$.
Given the optimal $\mathbf{x}^{*}$, we randomly sample $x_c$ percent of interactions from $\mathcal{D}_{u,c}$ for each category $c$ to construct the filtered interaction set $\mathcal{D}_u^{\text{filtered}}$.

This strategy intelligently allocates limited interaction resources, maximizing recommendation performance while ensuring a balanced representation of categories.

\subsubsection{Attention-aware Filtering}
This strategy uses Multi-Head Attention to find important user interactions. It selects useful interactions for better recommendations and efficient unlearning.

\textbf{Multi-Head Attention:} It uses $H$ attention heads to capture different aspects of information. 
For user $u$' interactions $\mathcal{D}_u = \{i_1, i_2, ..., i_m\}$ and candidate item $i_c$, each head $h \in \{1, 2, ..., H\}$ calculates an attention score:
\begin{equation}
\text{Attention}_h(i_c, \mathcal{D}_u) = \text{softmax}\left(\frac{\mathbf{Q}_h(i_c) \mathbf{K}_h(\mathcal{D}_u)^T}{\sqrt{d_k}}\right) \mathbf{V}_h(\mathcal{D}_u),
\end{equation}
where $\mathbf{Q}_h(i_c)$ is the query for $i_c$ in the head $h$. $\mathbf{K}_h(\mathcal{D}_u)$ and $\mathbf{V}_h(\mathcal{D}_u)$ are the keys and values for $\mathcal{D}_u$ in head $h$. $d_k$ is the key dimension.
The outputs of all $H$ heads are concatenated and then projected to the desired output dimension $d_{model}$ using a linear transformation with the output projection matrix $\mathbf{W}^O \in \mathbb{R}^{(H \times d_v) \times d_{model}}$:
\begin{equation}
\text{MultiHead}(i_c, \mathcal{D}_u) = \text{Concat}(\text{Attention}_1, ..., \text{Attention}_H) \mathbf{W}^O.
\end{equation}
\textbf{Attention Weight Calculation:} For candidate item $i_c$, it computes each interaction $i_j \in \mathcal{D}_u$'s weight $\alpha_{j,c}$:
\begin{equation}
\alpha_{j,c} = \frac{\exp(\text{score}(i_j, i_c))}{\sum_{k=1}^{m} \exp(\text{score}(i_k, i_c))}.
\end{equation}

Here, $\text{score}(i_j, i_c)$ can be from $\text{MultiHead}(i_c, \mathcal{D}_u)$. $\alpha_{j,c}$ shows how important $i_j$ is for $i_c$, and $m$ represents the number of user.

\textbf{Model-based Filtering:} We use the computed attention weights $\alpha_{j,c}$ to select the $K$ most relevant interactions from $\mathcal{D}_u$ for each candidate item $i_c$. We rank the interactions in $\mathcal{D}_u$ based on their attention weights $\alpha_{j,c}$ and select the Top-$K$ interactions to form the filtered interaction set $\mathcal{D}_u^{\text{filtered}}(i_c)$:
\begin{equation}
\mathcal{D}_u^{\text{filtered}}(i_c) = \text{Top-}K(\mathcal{D}_u, \{\alpha_{j,c}\}_{j=1}^m),
\end{equation}
where $\text{Top-}K(\mathcal{S}, \mathbf{w})$ is a function that returns the $K$ elements from set $\mathcal{S}$ with the highest weights in $\mathbf{w}$. 

\subsection{Prompt Construction and Augmented}
We make a prompt for the LLM. This prompt uses filtered user data, candidate items, and other information.
We use a template $P$ for the prompt:
\begin{equation}
P(u) = \operatorname{Format}(\mathcal{I}_u^{\text{cand}}, \mathcal{D}_u^{\text{filtered}}, C).
\end{equation}
Here, $\mathcal{I}_u^{\text{cand}}$ is the candidate items for user $u$ from the basic recommendation model. $\mathcal{D}_u^{\text{filtered}}$ is the filtered user data obtained in Section~\ref{Retrieval}. $C$ is extra information for the LLM, like user profile. $\operatorname{Format}$ puts these things into a prompt that the LLM can understand.

As shown in Appendix A.1, the prompt template $P(u)$ is designed to incorporate essential information for the LLM to make informed recommendations. The template includes the user's filtered interaction history ($\mathcal{D}_u^{\text{filtered}}$), which has been processed by our unlearning strategies to remove any unwanted information. This ensures that the LLM's recommendations are not influenced by data that needs to be forgotten. Additionally, the template includes candidate items ($\mathcal{I}_u^{\text{cand}}$) generated by a base recommendation model, providing a focused set of items for the LLM to consider. 
This will preserve the collaborative information of this user for user-level recommendation.
At the same time, we can also add historical interactions of similar users in the prompt to further strengthen the collaborative information and obtain better recommendation results.
We also incorporate auxiliary information ($C$), such as user profiles or item metadata, to enrich the context and enable the LLM to generate more accurate and personalized recommendations. The $\operatorname{Format}$ function structures these elements into a coherent prompt that the LLM can effectively interpret.

\subsection{LLM-based Recommendation Generation}
By using the prompt $P(u)$ constructed for each user, LLM is called to generate the result of the final recommendation.

$$
\hat{y}_u = f_{\text{LLM}}(P(u); \; \theta_{\text{LLM}}),
$$
where $\hat{y}_u$ is the generated recommendation for user $u$, $f_{\text{LLM}}$ represents a large language model that generates text, such as Llama or GPT-4 series, and $\theta_{\text{LLM}}$ represents the LLM's parameters.
In this paper, our model uses llama3.1-8b as the recommendation generator. 
Due to filtering in the retrieval phase, the LLM does not receive any information that needs to be forgotten, ensuring that the generated recommendation $\hat{y}_u$ complies with the requirements for forgetting.

\subsection{Privacy Analysis}

Our proposed method ensures precise unlearning by explicitly filtering out data to be unlearned during the retrieval phase, effectively preventing the LLM from accessing sensitive information. Consequently, this design inherently guarantees strong privacy protection.

Specifically, as the data intended for unlearning ($\mathcal{D}_u^{\text{unlearn}}$) are excluded during retrieval, recommendation score ($\hat{y}_u$) generated by the LLM solely rely on the filtered dataset ($\mathcal{D}_u^{\text{filtered}}$). This mechanism enforces conditional independence between the recommendation results and the sensitive, unlearned data, effectively preventing unintended leakage and strengthening privacy guarantees, aligning with established privacy-preserving practices in retrieval-augmented systems.

\section{EXPERIMENTS}
In this section, we evaluate the performance of CRAGRU by answering the following four research questions (RQs):
\begin{itemize}
    \item \textbf{RQ1:}  Can our method achieve performance comparable to retrained models and State-of-the-art (SOTA) models?

    \item \textbf{RQ2:}  What is CRAGRU’s time efficiency relative to existing exact and approximate unlearning approaches?

    \item \textbf{RQ3:} How effectively does CRAGRU eliminate the influence of forgotten data to ensure unlearning completeness?

    \item \textbf{RQ4:}  How do different retrieval strategies impact unlearning completeness and mitigate bias without degrading model utility?

\end{itemize}

\subsection{Experimental Setup}
\subsubsection{Datasets}
We evaluated our model on three publicly available datasets, which have been widely used for evaluating the performance of recommendation models. 
\textbf{\textit{i)} MovieLens 100K (ML-100K)}\footnote{https://grouplens.org/datasets/movielens/}: the MovieLens dataset is one of the most widely used datasets in recommendation research~\cite{harper2015movielens}, containing 100,000 user ratings. \textbf{\textit{ii)} MovieLens 1M (ML-1M)}: This is an extended version of the MovieLens dataset, containing 1,000,000 user ratings. \textbf{\textit{iii)} Netflix}\footnote{http://netflixprize.com}: This is the official dataset from the Netflix Prize competition.
Specifically, we reserve 10\% of the original dataset as the forgetting set $D_f$ and split the remaining data into training, validation, and test sets with a 70/10/20\% ratio. ~\autoref{tab:dataset_stats} provides a summary of the statistics for the three datasets, and Avg. Inter. represents the average number of user interactions in each dataset.
\begin{table}[ht]
\caption{Statistics of three public datasets.}
\label{tab:dataset_stats}
\centering
\begin{tabular}{lccccc}
\toprule
Dataset & Users & Items & Interactions & Avg. Inter.  & Sparsity \\ \midrule
ML-100K & 944 & 1,683 & 100,000 & 106 & 93.71\% \\
ML-1M & 6,041 & 3,707 & 1,000,209 & 165 & 95.53\% \\
Netflix & 7315 & 17,129 & 2,266,452 & 309 & 98.19\% \\
\bottomrule
\end{tabular}
\end{table}

\begin{table*}[!ht]
\centering
\caption{Comparison of different unlearning methods in terms of model utility, the best results are highlighted in bold. We computed the paired t-test p-values for CRAGRU and RecEraser on three datasets, and all p-values were less than 0.01.}
\label{tab:RQ1}
\resizebox{0.97\textwidth}{!}{
\begin{tabular}{ccccccccccccc}
\toprule
\multirow{2}{*}{\textbf{ML-100K}} & \multicolumn{6}{c}{\textbf{BPR}} & \multicolumn{6}{c}{\textbf{LightGCN}} \\
\cmidrule(lr){2-7} \cmidrule(lr){8-13}
 & \textbf{HR@5} & \textbf{NDCG@5} & \textbf{HR@10} & \textbf{NDCG@10} & \textbf{HR@20} & \textbf{NDCG@20} & \textbf{HR@5} & \textbf{NDCG@5} & \textbf{HR@10} & \textbf{NDCG@10} & \textbf{HR@20} & \textbf{NDCG@20}  \\

\midrule
Retrain & 0.6643&0.2914&0.7845&0.2910&0.8728&0.3091&0.6455&0.2796&0.7739&0.2790&0.8634&0.2963 \\
SISA & 0.2968&0.0908&0.4240&0.0949&0.5607&0.1085&0.2733&0.0818&0.3899&0.0849&0.5183&0.0925 \\
GraphEraser & 0.3239&0.1011&0.4488&0.1093&0.6148&0.1282&0.3805&0.1156&0.5336&0.1257&0.6808&0.1460 \\
RecEraser & 0.3204&0.1011&0.4582&0.1100&0.6337&0.1274&0.4806&0.1557&0.6137&0.1668&0.7491&0.1873 \\
SCIF & 0.2524&0.0720&0.4093&0.0851&0.6055&0.1074&0.4857&0.1736&0.6098&0.1707&\textbf{0.7370}&0.1840 \\
IFRU & 0.3030&0.1899&0.4747&0.2432&0.5858&0.2710&0.2501&0.1629&0.4092&0.2138&0.6258&\textbf{0.2685} \\
\textbf{CRAGRU} &\textbf{0.6193}&\textbf{0.2864}&\textbf{0.7116}&\textbf{0.2801}&\textbf{0.7805}&\textbf{0.2886}&\textbf{0.5366}&\textbf{0.2241}&\textbf{0.6384}&\textbf{0.2156}&0.7169&0.2216 \\
\toprule
\multirow{2}{*}{\textbf{ML-1M}} & \multicolumn{6}{c}{\textbf{BPR}} & \multicolumn{6}{c}{\textbf{LightGCN}} \\
\cmidrule(lr){2-7} \cmidrule(lr){8-13}
 & \textbf{HR@5} & \textbf{NDCG@5} & \textbf{HR@10} & \textbf{NDCG@10} & \textbf{HR@20} & \textbf{NDCG@20} & \textbf{HR@5} & \textbf{NDCG@5} & \textbf{HR@10} & \textbf{NDCG@10} & \textbf{HR@20} & \textbf{NDCG@20}  \\

\midrule
Retrain & 0.6185&0.2673&0.7403&0.2555&0.8352&0.2598&0.6111&0.2670&0.7377&0.2533&0.8339&0.2582 \\
SISA & 0.2432&0.0731&0.3547&0.0707&0.4755&0.0738&0.2093&0.0593&0.2964&0.0588&0.3928&0.0624 \\
GraphEraser & 0.3137&0.0898&0.4540&0.0902&0.6126&0.0995&0.4249&0.1283&0.5635&0.1295&0.6968&0.1388 \\
RecEraser & 0.3514&0.1084&0.4833&0.1055&0.6260&0.1123&0.4560&0.1464&0.6003&0.1467&0.7379&0.1574 \\
SCIF & 0.3944&0.1338&0.5341&0.1332&0.6833&0.1436&0.4823&0.1796&0.6060&0.1723&0.7230&0.1764 \\
IFRU & 0.2820&0.1796&0.4227&\textbf{0.2248}&0.5749&\textbf{0.2634}&0.2656&0.1659&0.4350&0.2205&0.6502&0.2746 \\
\textbf{CRAGRU} & \textbf{0.5377}&\textbf{0.2247}&\textbf{0.6412}&0.2127&\textbf{0.7227}&0.2167&\textbf{0.5520}&\textbf{0.2354}&\textbf{0.6556}&\textbf{0.2221}&\textbf{0.7406}&\textbf{0.2270} \\
\toprule
\multirow{2}{*}{\textbf{Netflix}} & \multicolumn{6}{c}{\textbf{BPR}} & \multicolumn{6}{c}{\textbf{LightGCN}} \\
\cmidrule(lr){2-7} \cmidrule(lr){8-13}
 & \textbf{HR@5} & \textbf{NDCG@5} & \textbf{HR@10} & \textbf{NDCG@10} & \textbf{HR@20} & \textbf{NDCG@20} & \textbf{HR@5} & \textbf{NDCG@5} & \textbf{HR@10} & \textbf{NDCG@10} & \textbf{HR@20} & \textbf{NDCG@20}  \\

\midrule
Retrain &0.7927&0.3382&0.9041&0.2995&0.9564&0.2555&0.7725&0.3329&0.8879&0.2928&0.9523&0.2485\\
SISA & 0.1167&0.0248&0.2201&0.0256&0.3645&0.0252&0.1816&0.0487&0.2635&0.0471&0.3603&0.0439 \\
GraphEraser & 0.3980&0.0941&0.6343&0.1001&0.8099&0.0957&0.4878&0.1157&0.7032&0.1215&0.8618&0.1181 \\
RecEraser & 0.4303&0.1032&0.6673&0.1086&0.8361&0.1038&0.4959&0.1303&0.7129&0.1314&\textbf{0.8772}&0.1254 \\
SCIF & 0.2789&0.0793&0.4014&0.0734&0.5265&0.08270&0.3388&0.1002&0.4762&0.0931&0.6286&0.1042 \\
IFRU & 0.1111&0.0677&0.2020&0.0955&0.3030&0.1202&0.1818&0.1035&0.2600&0.1388&0.3508&0.1760 \\
\textbf{CRAGRU} & \textbf{0.7266}&\textbf{0.3230}&\textbf{0.8190}&\textbf{0.2759}&\textbf{0.8651}&\textbf{0.2286}&\textbf{0.6691}&\textbf{0.2717}&\textbf{0.7827}&\textbf{0.2400}&0.8428&\textbf{0.2072}\\
\bottomrule
\end{tabular}}

\end{table*}

\subsubsection{Compared Models}
Our method is model-agnostic and can be applied to any recommendation model. In this paper, we compare CRAGRU with State-of-the-art unlearning methods across two representative recommendation models:
\begin{itemize}
    \item \textbf{BPR}~\cite{rendle2012bpr}: This is a widely used recommendation model, where the core idea is to optimize matrix factorization using a Bayesian personalized ranking objective.
    \item \textbf{LightGCN}~\cite{he2020lightgcn}: This is an advanced collaborative filtering model that improves recommendation performance by simplifying the graph convolutional network.
\end{itemize}

\noindent The compared unlearning methods are listed as follows:
\begin{itemize}
    \item \textbf{Retrain}: Retrain achieves unlearning by retraining the model in the dataset after removing the forget set.
    \item \textbf{SISA}~\cite{bourtoule2021machine}: This is an unlearning method that divides the data into multiple shards, trains sub-models independently on each shard, and aggregates their prediction results.
    \item \textbf{GraphEraser}~\cite{chen2022graph}: This is an unlearning method designed for graph data. It partitions the graph through node clustering and uses static weighted aggregation for prediction to better align with the characteristics of graph data.
    \item \textbf{RecEraser}~\cite{chen2022recommendation}: It is an advanced recommendation unlearning method that improves the partitioning and aggregation strategies of SISA by using interaction-based partitioning and attention-based aggregation to improve recommendation performance.
    \item \textbf{SCIF}~\cite{li2023selective}: SCIF uses influence functions to negate removed data’s impact but ignores spillover effects. It never truly deletes records; it replaces labels with their average.
    \item \textbf{IFRU}~\cite{zhang2024recommendation}: IFRU iteratively reweights affected latent features to remove forgotten interactions, but this process can be computationally intensive and may fail to fully isolate the forget set’s impact.

\end{itemize}

\subsubsection{Evaluation Metrics}
To evaluate recommendation performance after unlearning, we employ two common metrics: Hit Ratio (HR@$K$) and Normalized Discounted Cumulative Gain (NDCG@$K$). HR@$K$ measures the precision of the recommendation by calculating the proportion of times that the user's target item is present in the top$K$ recommendations. NDCG@$K$ is a ranking metric that gives higher weights to top-ranked items, considering their positions in the recommendation list. Both metrics are calculated on the remaining interaction data $D_r$ after unlearning. In our experiments, we evaluate the performance at $K \in \{5, 10, 20\}$. Higher values indicate better performance. Furthermore, we measure unlearning efficiency by comparing the unlearning time; a shorter unlearning time indicates greater efficiency.

\subsection{Model Utility (RQ1)}
\label{sec:RQ1}

We evaluate the model utility of CRAGRU on three public datasets using HR@K and NDCG@K (K = 5, 10, 20) under two backbone models: BPR and LightGCN. To simulate unlearning requests, 10\% of user interactions are randomly selected and removed. This experiment not only measures the absolute performance of each method, but more importantly, reflects the extent to which forgetting targeted users negatively affects the recommendation utility for others—i.e., unlearning bias.

We compare CRAGRU against six baselines, including retraining, partition-based methods (SISA, GraphEraser, RecEraser), and approximate unlearning methods (SCIF, IFRU). Retraining achieves the highest performance since no actual unlearning is required, but at impractical computational cost. Among partition-based methods, RecEraser performs best, as it aggregates similar users to reduce utility loss. However, these methods often suffer from unlearning bias due to the entanglement of forgotten users with co-located users in training shards. Approximate methods (SCIF, IFRU) reduce computation by estimating user influence via gradients or similarity propagation, but can cause latent drift in similar users’ embeddings, leading to degraded recommendation quality.

CRAGRU outperforms almost all other unlearning baselines across datasets and backbones while approaching retraining performance, indicating strong bias mitigation. For instance, on the ML-1M dataset with LightGCN, CRAGRU improves HR@10 by 9.64\% and NDCG@10 by 12.3\% over RecEraser. These gains reflect CRAGRU’s ability to precisely isolate and remove target user influence at the retrieval stage, thus minimizing adverse impacts on behaviorally similar users. Performance improvements are statistically significant ($p<0.01$) across all metrics and datasets.
This experiment confirms that CRAGRU achieves near-retraining performance while avoiding the performance degradation and bias propagation common in both exact and approximate unlearning baselines.

\subsection{Unlearning Efficiency (RQ2)}
Our method can achieve unlearning in the LLM inference phase,  which is equivalent to the training process of other methods. Therefore, when evaluating the unlearning efficiency of different methods, we ignore the inference time of other models and evaluate by comparing the training time required for each method to complete the same unlearning task. The experimental results demonstrate that our method has a significant time advantage.
Specifically, we randomly selected a user from the dataset, and the average interactions of the user between different datasets are shown in \autoref{tab:dataset_stats}. Our task was to perform unlearning on all interactions of this user. For partition-aggregation-based methods, i.e. SISA, GraphEraser, and RecEraser, we followed their recommended parameter settings. All experiments were run on an NVIDIA GeForce RTX 4090 GPU and the total unlearning time was recorded for each method. The results are shown in \autoref{tab:training_time}.
The lower the time, the higher the unlearning efficiency. Bold text in the table represents the best values, while underlined text represents the second-best values. Our method achieved the best results across all datasets.
\begin{table}[ht]
\centering
\caption{Training Time Comparison on Different Datasets and Models}
\label{tab:training_time}
\vspace{-6pt}
\resizebox{0.47\textwidth}{!}{
\begin{tabular}{lcccccc}
\toprule
 & \multicolumn{2}{c}{ML-100K} & \multicolumn{2}{c}{ML-1M} & \multicolumn{2}{c}{Netflix} \\
\cmidrule(lr){2-3} \cmidrule(lr){4-5} \cmidrule(lr){6-7}
 & BPR & LightGCN & BPR & LightGCN & BPR & LightGCN \\
\midrule
Retrain & 248s & 141s & 1935s & 4209s & 527s & 1019s \\
SISA & 27s & 28s & 298s & 435s & \underline{116s} & 131s \\
GraphEraser & \underline{26s} & \underline{17s} & 302s & 269s & 310s & 640s \\
RecEraser & 29s & 35s & 386s & 275s & 404s & 880s \\
SCIF & 18s & 18s &  \underline{66s} & \underline{64s} & 299s & 177s \\
IFRU & 55s & 57s &  78s & 90s & 104s & \underline{117s} \\
\textbf{CRAGRU} & \textbf{14s} & \textbf{14s} & \textbf{15s} & \textbf{15s} & \textbf{16s} & \textbf{17s} \\
\bottomrule
Improve & {1.8x} & 1.2x & 4.4x & 4.3x & 7.3x & 6.9x \\
\bottomrule
\end{tabular}}
\end{table}
As shown in \autoref{tab:training_time}, compared to Retrain, the partition-aggregation framework (i.e., SISA, GraphEraser, and RecEraser) significantly improved the unlearning efficiency.
Moreover, since SISA uses a very simple partition and aggregation strategy, it is faster than the other two partition aggregation frameworks. However, this simplicity also limits the performance of SISA in terms of model utility (see \autoref{tab:RQ1}).
Our method demonstrates a significant advantage in unlearning time efficiency. Compared to the second-best values of these methods, it achieved average speedups of 1.5x, 4.4x, and 7.1x in the three datasets, respectively.
This is because we leverage LLMs to shift the recommendation task from the dataset level to the user level. This atomic user-level recommendation enables more flexible control over each user's recommendations and unlearning, thereby enhancing unlearning efficiency. 
As the number of unlearning interactions increases, the time consumption of our method grows linearly, while partition-aggregation methods experience exponential growth. \textit{According to Liu et al.~\cite{liu2022forgetting}, when the dataset is split into 10 partitions and 100 interactions are randomly selected for unlearning, the probability that all sub-models require retraining is close to 100\%.} 
In real-world scenarios, user unlearning requests typically come gradually, so the user-level unlearning method can quickly fulfill these needs in a timely manner.
\begin{figure*}[ht]
  \centering
  \vspace{-15pt}
  \includegraphics[width=1\textwidth]{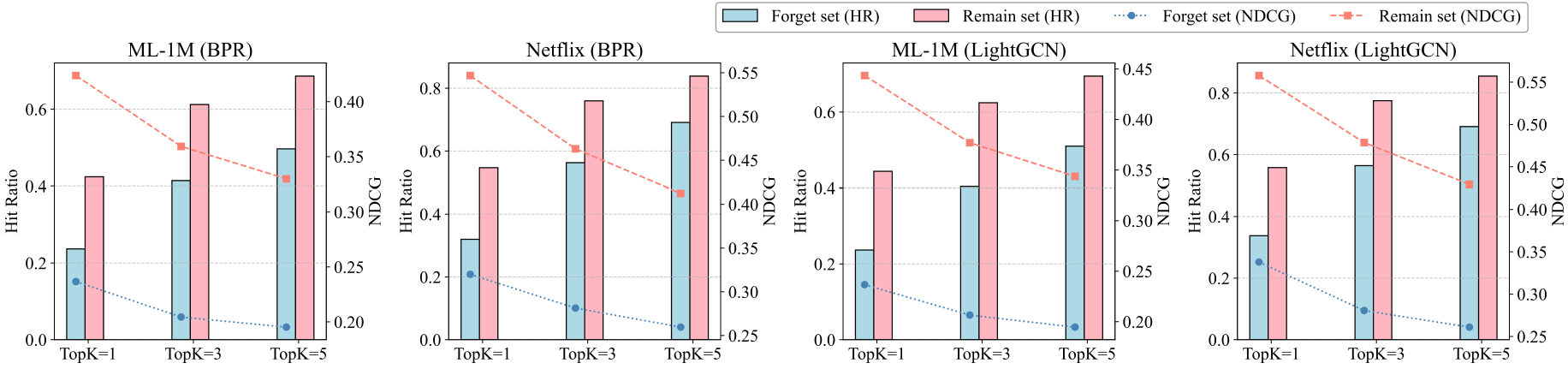}\
  \vspace{-20pt}
  \caption{Comparison of the performance between the forget set and the remain set on ML-1M and Netflix datasets.}
  \label{exp3}
\end{figure*}
\begin{figure*}[ht]
  \centering
  \vspace{-10pt}
  \includegraphics[width=1\textwidth]{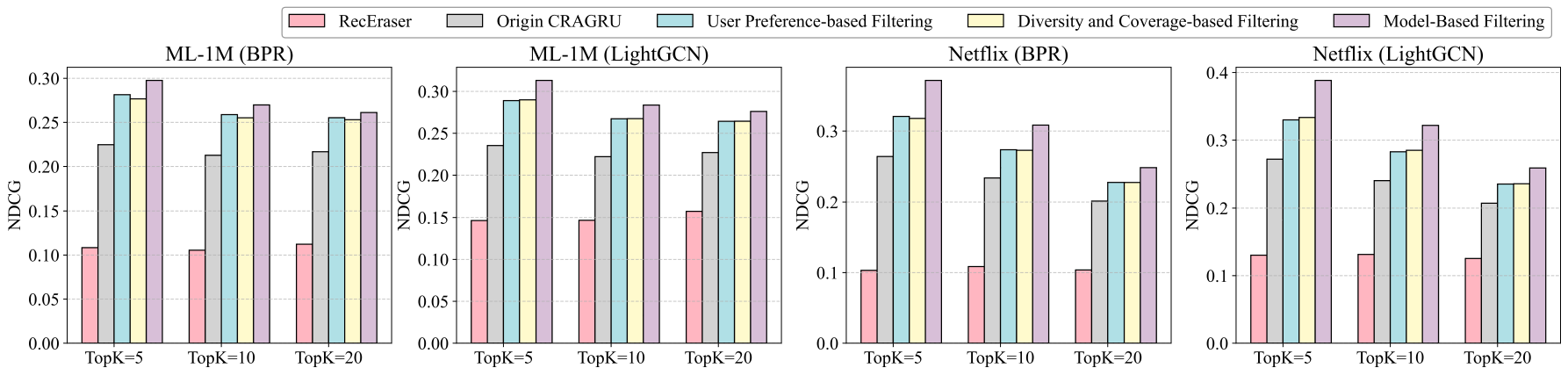}
  \vspace{-20pt}
  \caption{Comparison of the performance of different retrieval strategies for the CRAGRU model on the ML-1M and Netflix.}
  \vspace{-14pt}
  \label{exp4_NDCG}
\end{figure*}

\subsection{Unlearning Completeness (RQ3)}
To assess the effectiveness of CRAGRU in erasing user-specific influence, we compare recommendation performance on the \textit{forgotten set} versus the \textit{remaining set}, following simulated unlearning of 10\% user-item interactions on ML-1M and Netflix datasets. BPR and LightGCN are used as backbone models. We report HR@K and NDCG@K for $K=1,3,5$ to measure personalized recommendation quality.

Unlike traditional unlearning methods that often leave residual influence of forgotten users on the model, CRAGRU filters these users’ traces at the retrieval stage, ensuring atomic removal. As shown in \autoref{exp3}, the recommendation quality for the forgotten set is consistently and significantly lower than that of the remaining set across all settings, indicating effective removal of memorized patterns and minimal cross-user leakage.
For instance, on ML-1M with BPR, the HR@1 and NDCG@1 of the forgotten set are only 55.85\% of those of the remaining set; As $K$ increases to 3 and 5, the HR and NDCG ratios rise to 67.61\% / 56.89\% and 72.36\% / 59.19\%, respectively. Similar trends are observed for Netflix and LightGCN.

This consistent performance drop in the forgotten set demonstrates that CRAGRU effectively localizes the unlearning effect to the target users without impacting the recommendation quality for others. Moreover, as $K$ increases, the performance gap narrows—this is likely due to the expansion of the candidate item pool $\mathcal{I}_u^{\text{cand}}$ provided by the backbone model, introducing items with weaker user relevance and reducing filtering precision.
In summary, CRAGRU achieves high unlearning effectiveness while preserving the performance of non-target users, addressing the central challenge of unlearning bias in recommendation systems.

\subsection{Effectiveness of Retrieval Strategies (RQ4)}

This experiment investigates how CRAGRU’s three retrieval filtering strategies contribute to both recommendation quality and unlearning bias mitigation. We compare them with the original backbone model (without retrieval filtering) and the other unlearning method RecEraser on ML-1M and Netflix datasets, using BPR and LightGCN as backbones. \autoref{exp4_NDCG} reports NDCG@K for $K=5,10,20$.

CRAGRU consistently outperforms RecEraser across datasets and backbones, indicating superior preservation of recommendation quality while performing user-level unlearning. For example, on ML-1M with BPR, CRAGRU improves NDCG@10 by 10.72\% over RecEraser; with LightGCN, the gain is 7.54\%.

All three retrieval strategies improve over the unfiltered CRAGRU baseline, demonstrating that retrieval-stage filtering is critical to reducing collateral performance loss—a direct manifestation of unlearning bias. 
1) User preference-based filtering performs well by preserving long-term semantic consistency, e.g., on Netflix (LightGCN), improving NDCG@10 from 0.2400 to 0.2827. 2) Diversity-aware retrieval ensures representation of different item clusters to avoid overfitting to specific behaviors. While slightly less precise than preference-based filtering, it reduces retrieval bias with lower computational cost via global distribution optimization.
3) Attention-aware retrieval—using multi-head attention to score and prioritize impactful interactions—achieves the highest gains. On ML-1M, it improves NDCG@10 by 5.71\% (BPR) and 6.16\% (LightGCN). This is because it can capture complex relationships between user interactions and candidate items more effectively by evaluating the importance of interaction from multiple perspectives.
Compared with approximate unlearning methods such as SCIF and IFRU (see \autoref{tab:RQ1}), which often suffer from residual embedding shifts, CRAGRU’s filtering strategies ensure that only unbiased, non-forgotten interactions contribute to the prompt, avoiding semantic leakage.

\section{Conclusion}
In this paper, we address a key limitation of existing recommendation unlearning methods—their tendency to propagate unintended influence to related users due to coarse-grained unlearning at the dataset level. To overcome this, we propose CRAGRU, a novel framework that leverages Retrieval-Augmented Generation (RAG) to enable fine-grained, user-level unlearning while minimizing unlearning bias. CRAGRU employs three retrieval filtering strategies to isolate the impact of forgotten users and preserve recommendation quality for non-target users. Extensive experiments on three real-world datasets with BPR and LightGCN backbones demonstrate that CRAGRU outperforms state-of-the-art models in both efficiency and utility. 

\bibliographystyle{elsarticle-num} 
\bibliography{icdm}

\begin{thebibliography}{10}
\expandafter\ifx\csname url\endcsname\relax
  \def\url#1{\texttt{#1}}\fi
\expandafter\ifx\csname urlprefix\endcsname\relax\def\urlprefix{URL }\fi
\expandafter\ifx\csname href\endcsname\relax
  \def\href#1#2{#2} \def\path#1{#1}\fi

\bibitem{he2020lightgcn}
X.~He, K.~Deng, X.~Wang, Y.~Li, Y.~Zhang, M.~Wang, Lightgcn: Simplifying and powering graph convolution network for recommendation, in: ACM SIGIR 2020, 2020, pp. 639--648.

\bibitem{liu2023federated}
W.~Liu, C.~Chen, X.~Liao, M.~Hu, J.~Yin, Y.~Tan, L.~Zheng, Federated probabilistic preference distribution modelling with compactness co-clustering for privacy-preserving multi-domain recommendation., in: IJCAI, 2023, pp. 2206--2214.

\bibitem{hu2008collaborative}
Y.~Hu, Y.~Koren, C.~Volinsky, Collaborative filtering for implicit feedback datasets, in: ICDM 2008, Ieee, 2008, pp. 263--272.

\bibitem{gdpr2016general}
G.~GDPR, General data protection regulation, Regulation (EU) 679 (2016).

\bibitem{huang2021data}
H.~Huang, J.~Mu, N.~Z. Gong, Q.~Li, B.~Liu, M.~Xu, Data poisoning attacks to deep learning based recommender systems, arXiv preprint arXiv:2101.02644 (2021).

\bibitem{wang2024comprehensive}
Z.~Wang, E.~Yang, L.~Shen, H.~Huang, A comprehensive survey of forgetting in deep learning beyond continual learning, IEEE TPAMI (2024).

\bibitem{chen2021machine}
M.~Chen, Z.~Zhang, T.~Wang, M.~Backes, M.~Humbert, Y.~Zhang, When machine unlearning jeopardizes privacy, in: ACM SIGSAC, 2021, pp. 896--911.

\bibitem{sachdeva2024machine}
B.~Sachdeva, H.~Rathee, Sristi, A.~Sharma, W.~Wydma{\'n}ski, Machine unlearning for recommendation systems: An insight, in: International Conference On Innovative Computing And Communication, Springer, 2024, pp. 415--430.

\bibitem{chen2022graph}
M.~Chen, Z.~Zhang, T.~Wang, M.~Backes, M.~Humbert, Y.~Zhang, Graph unlearning, in: 2022 ACM SIGSAC, 2022, pp. 499--513.

\bibitem{chen2022recommendation}
C.~Chen, F.~Sun, M.~Zhang, B.~Ding, Recommendation unlearning, in: ACM Web Conference, 2022, pp. 2768--2777.

\bibitem{zhang2024recommendation}
Y.~Zhang, Z.~Hu, Y.~Bai, J.~Wu, Q.~Wang, F.~Feng, Recommendation unlearning via influence function, ACM TORS 3~(2) (2024) 1--23.

\bibitem{chen2024post}
C.~Chen, Y.~Zhang, Y.~Li, J.~Wang, L.~Qi, X.~Xu, X.~Zheng, J.~Yin, Post-training attribute unlearning in recommender systems, ACM TOIS 43~(1) (2024) 1--28.

\bibitem{li2023selective}
Y.~Li, C.~Chen, X.~Zheng, Y.~Zhang, B.~Gong, J.~Wang, L.~Chen, Selective and collaborative influence function for efficient recommendation unlearning, ESWA 234 (2023) 121025.

\bibitem{koh2017understanding}
P.~W. Koh, P.~Liang, Understanding black-box predictions via influence functions, in: TCML, PMLR, 2017, pp. 1885--1894.

\bibitem{li2024survey}
Y.~Li, X.~Feng, C.~Chen, Q.~Yang, A survey on recommendation unlearning: Fundamentals, taxonomy, evaluation, and open questions, arXiv preprint arXiv:2412.12836 (2024).

\bibitem{li2025causal}
M.~Li, H.~Sui, Causal recommendation via machine unlearning with a few unbiased data, in: AAAI 2025 Workshop on Artificial Intelligence with Causal Techniques, 2025, p.~.

\bibitem{liu2022forgetting}
W.~Liu, J.~Wan, X.~Wang, W.~Zhang, D.~Zhang, H.~Li, Forgetting fast in recommender systems, arXiv preprint arXiv:2208.06875 (2022).

\bibitem{nguyen2022survey}
T.~T. Nguyen, T.~T. Huynh, Z.~Ren, P.~L. Nguyen, A.~W.-C. Liew, H.~Yin, Q.~V.~H. Nguyen, A survey of machine unlearning, arXiv preprint arXiv:2209.02299 (2022).

\bibitem{tarun2023deep}
A.~K. Tarun, V.~S. Chundawat, M.~Mandal, M.~Kankanhalli, Deep regression unlearning, in: ICLR, PMLR, 2023, pp. 33921--33939.

\bibitem{brophy2021machine}
J.~Brophy, D.~Lowd, Machine unlearning for random forests, in: ICLR, PMLR, 2021, pp. 1092--1104.

\bibitem{ginart2019making}
A.~Ginart, M.~Guan, G.~Valiant, J.~Y. Zou, Making ai forget you: Data deletion in machine learning, NeurIPS 32 (2019).

\bibitem{cauwenberghs2000incremental}
G.~Cauwenberghs, T.~Poggio, Incremental and decremental support vector machine learning, NeurIPS 13 (2000).

\bibitem{karasuyama2010multiple}
M.~Karasuyama, I.~Takeuchi, Multiple incremental decremental learning of support vector machines, IEEE TNN 21~(7) (2010) 1048--1059.

\bibitem{cao2015towards}
Y.~Cao, J.~Yang, Towards making systems forget with machine unlearning, in: 2015 IEEE SP, IEEE, 2015, pp. 463--480.

\bibitem{bourtoule2021machine}
L.~Bourtoule, V.~Chandrasekaran, C.~A. Choquette-Choo, H.~Jia, A.~Travers, B.~Zhang, D.~Lie, N.~Papernot, Machine unlearning, in: 2021 IEEE SP, IEEE, 2021, pp. 141--159.

\bibitem{chen2023unlearn}
J.~Chen, D.~Yang, Unlearn what you want to forget: Efficient unlearning for llms, arXiv preprint arXiv:2310.20150 (2023).

\bibitem{yao2023large}
Y.~Yao, X.~Xu, Y.~Liu, Large language model unlearning, arXiv preprint arXiv:2310.10683 (2023).

\bibitem{xu2023netflix}
M.~Xu, J.~Sun, X.~Yang, K.~Yao, C.~Wang, Netflix and forget: Efficient and exact machine unlearning from bi-linear recommendations, arXiv preprint arXiv:2302.06676 (2023).

\bibitem{yuan2023federated}
W.~Yuan, H.~Yin, F.~Wu, S.~Zhang, T.~He, H.~Wang, Federated unlearning for on-device recommendation, in: WSDM 2023, 2023, pp. 393--401.

\bibitem{xin2024effectiveness}
X.~Xin, L.~Yang, Z.~Zhao, P.~Ren, Z.~Chen, J.~Ma, Z.~Ren, On the effectiveness of unlearning in session-based recommendation, in: WSDM 2024, 2024, pp. 855--863.

\bibitem{ye2023sequence}
S.~Ye, J.~Lu, Sequence unlearning for sequential recommender systems, in: Australasian Joint Conference on Artificial Intelligence, Springer, 2023, pp. 403--415.

\bibitem{zhao2023recommender}
Z.~Zhao, W.~Fan, J.~Li, Y.~Liu, X.~Mei, Y.~Wang, Z.~Wen, F.~Wang, X.~Zhao, J.~Tang, et~al., Recommender systems in the era of large language models (llms), arXiv preprint arXiv:2307.02046 (2023).

\bibitem{lin2023can}
J.~Lin, X.~Dai, Y.~Xi, W.~Liu, B.~Chen, H.~Zhang, Y.~Liu, C.~Wu, X.~Li, C.~Zhu, et~al., How can recommender systems benefit from large language models: A survey, arXiv preprint arXiv:2306.05817 (2023).

\bibitem{liu2023pre}
P.~Liu, L.~Zhang, J.~A. Gulla, Pre-train, prompt, and recommendation: A comprehensive survey of language modeling paradigm adaptations in recommender systems, TACL 11 (2023) 1553--1571.

\bibitem{wu2024survey}
L.~Wu, Z.~Zheng, Z.~Qiu, H.~Wang, H.~Gu, T.~Shen, C.~Qin, C.~Zhu, H.~Zhu, Q.~Liu, et~al., A survey on large language models for recommendation, World Wide Web 27~(5) (2024) 60.

\bibitem{shu2024knowledge}
D.~Shu, T.~Chen, M.~Jin, C.~Zhang, M.~Du, Y.~Zhang, Knowledge graph large language model (kg-llm) for link prediction, arXiv preprint arXiv:2403.07311 (2024).

\bibitem{geng2022recommendation}
S.~Geng, S.~Liu, Z.~Fu, Y.~Ge, Y.~Zhang, Recommendation as language processing (rlp): A unified pretrain, personalized prompt \& predict paradigm (p5), in: ACM RecSys, 2022, pp. 299--315.

\bibitem{cui2022m6}
Z.~Cui, J.~Ma, C.~Zhou, J.~Zhou, H.~Yang, M6-rec: Generative pretrained language models are open-ended recommender systems, arXiv preprint arXiv:2205.08084 (2022).

\bibitem{gao2023chat}
Y.~Gao, T.~Sheng, Y.~Xiang, Y.~Xiong, H.~Wang, J.~Zhang, Chat-rec: Towards interactive and explainable llms-augmented recommender system, arXiv preprint arXiv:2303.14524 (2023).

\bibitem{zhang2023recommendation}
J.~Zhang, R.~Xie, Y.~Hou, X.~Zhao, L.~Lin, J.-R. Wen, Recommendation as instruction following: A large language model empowered recommendation approach, ACM TOIS (2023).

\bibitem{bao2023tallrec}
K.~Bao, J.~Zhang, Y.~Zhang, W.~Wang, F.~Feng, X.~He, Tallrec: An effective and efficient tuning framework to align large language models with recommendations, in: RecSys 2023, 2023, pp. 1007--1014.

\bibitem{wei2024llmrec}
W.~Wei, X.~Ren, J.~Tang, Q.~Wang, L.~Su, S.~Cheng, J.~Wang, D.~Yin, C.~Huang, Llmrec: Large language models with graph augmentation for recommendation, in: WSDM 2024, 2024, pp. 806--815.

\bibitem{ren2024representation}
X.~Ren, W.~Wei, L.~Xia, L.~Su, S.~Cheng, J.~Wang, D.~Yin, C.~Huang, Representation learning with large language models for recommendation, in: WWW, 2024, pp. 3464--3475.

\bibitem{li2024making}
Y.~Li, C.~Chen, X.~Zheng, J.~Liu, J.~Wang, Making recommender systems forget: Learning and unlearning for erasable recommendation, Knowledge-Based Systems 283 (2024) 111124.

\bibitem{li2023ultrare}
Y.~Li, C.~Chen, Y.~Zhang, W.~Liu, L.~Lyu, X.~Zheng, D.~Meng, J.~Wang, Ultrare: Enhancing receraser for recommendation unlearning via error decomposition, Advances in Neural Information Processing Systems 36 (2023) 12611--12625.

\bibitem{harper2015movielens}
F.~M. Harper, J.~A. Konstan, The movielens datasets: History and context, TIIS 5~(4) (2015) 1--19.

\bibitem{rendle2012bpr}
S.~Rendle, C.~Freudenthaler, Z.~Gantner, L.~Schmidt-Thieme, Bpr: Bayesian personalized ranking from implicit feedback, arXiv preprint arXiv:1205.2618 (2012).

\end{thebibliography}

\newpage
\appendix
\section{Implementation Details}
\subsection{Prompts Template}
\label{subsec:prompt}
\begin{Verbatim}[frame=single, breaklines=true, breaksymbol={}, fontsize=\small]
I want you to predict the user‘s rating for each movie in the candidate list on a scale from 1 to 100, based on the user’s profile and movie interaction history. Follow these instructions carefully:
1. Use the given user profile and historical movie interaction records to predict how much the user would like each movie in the candidate list. The higher the score, the more likely the user will enjoy the movie.
2. The output must be in valid JSON format, where each movie ID is paired with its predicted score.  The format should be:
{ "movie_id1": score1, 
"movie_id2" : score2,
...}
3. Ensure that all movie IDs in the candidate list are included exactly once in the output.
4. Do not include any additional text, explanation, or comments outside the JSON object.
### User Profile:
{user_profile_text}.
### Movie Interaction History:
The user's historical movie interaction records include:{history_movies}
### Candidate List:
{candidate_list}
Predict and output the ratings in the required JSON format.
\end{Verbatim}
\vspace{5pt}

\begin{Verbatim}[frame=single, breaklines=true, breaksymbol={}, fontsize=\small]
I want you to predict the user‘s rating for each movie in the candidate list on a scale from 1 to 100,  based on the user’s interaction history. Follow these instructions carefully:
1. Use the given user‘s historical movie interaction records to predict how much the user would like each movie in the candidate list. The higher the score, the more likely the user will enjoy the movie.
2. The output must be in valid JSON format, where each movie ID is paired with its predicted score. 
The format should be:
{ "movie_id1" : score1,
  "movie_id2" : score2,
...}
3. Ensure that all movie IDs in the candidate list are included exactly once in the output.
4. Do not include any additional text, explanation, or comments outside the JSON object.
### Movie Interaction History:
The user's historical movie interaction records include:
{history_movies}
### Candidate List: {candidate_list}
Predict and output the ratings in the required JSON format.
\end{Verbatim}

This section presents the prompt templates used in our experiments, with distinct templates designed for the MovieLens and Netflix datasets to suit the characteristics of each dataset and the specific task requirements, as shown above.

In this, \textbf{\{user\_profile\_text\}} represents a natural language sentence composed of each user's personalized attributes, \textbf{\{history\_movies\}} is the list of movies the user has watched, with each movie including its title, genre, and release year, and \textbf{\{candidate\_list\}} is the top-K list of movies recommended to each user by traditional recommendation models.
\vspace{-0.1cm}
\subsection{Algorithm Implementation}

\begin{algorithm}[htb]
    \caption{CRAGRU Recommendation Unlearning Process}
    \label{alg:cragru}
    \begin{algorithmic}[1]
        \Require $\mathcal{D}$: User interaction data, $\mathcal{D}_f^u$: Unlearning data, $\mathcal{I}_u^{cand}$: Candidate items, $C$: Auxiliary information, $f_{\text{LLM}}$: LLM, $\theta_{\text{LLM}}$: LLM parameters, $K$: Number of interactions to retain.
        \Ensure Recommended items $\hat{y}_u$ for user $u$.
        \State \textbf{Retrieval:}
        \State $\mathcal{D}_u = \operatorname{Retrieve}(u; \mathcal{D})$ \Comment{Retrieve user $u$'s interaction data}
        \State $\mathcal{D}_u^{\text{filtered}} = \operatorname{Filter}(\mathcal{D}_u, \mathcal{D}_f^u)$ \Comment{Filter unlearning data}

        \State \textbf{Filter Function Details:}
        \State \textbf{Strategy 1: User Preference-based Filtering}
        \State $\mathcal{C} = \{c_1, c_2, ..., c_n\}$ \Comment{Define item categories}
        \State Calculate $p_{u,c}$ for each $c \in \mathcal{C}$ \Comment{Interaction proportion}
        \State Calculate $K_c = \lfloor p_{u,c} \times K \rfloor$ for each $c \in \mathcal{C}$
        \State $\mathcal{D}_u^{\text{filtered}} = \{ x \mid x \in \operatorname{Sample}(\mathcal{D}_{u,c}, K_c), \, c \in \mathcal{C} \}$

        \State \textbf{Strategy 2: Diversity and Coverage-based Filtering}
        \State Pre-compute performance matrix $\mathbf{M}$
        \State Solve knapsack problem to get optimal retention ratios $\mathbf{x}^*$
        \State $\mathcal{D}_u^{\text{filtered}} = \bigcup_{c \in \mathcal{C}} \operatorname{Sample}(\mathcal{D}_{u,c}, x_c \times |\mathcal{D}_{u,c}|)$

        \State \textbf{Strategy 3: Model-Based Filtering}
        \State Calculate attention weights $\alpha_{j,c}$ for each interaction $i_j \in \mathcal{D}_u$ and candidate item $i_c$
        \State $\mathcal{D}_u^{\text{filtered}}(i_c) = \text{Top-}K(\mathcal{D}_u, \{\alpha_{j,c}\}_{j=1}^m)$  \Comment{Select top K interactions for each candidate}
        \State $\mathcal{D}_u^{\text{filtered}} = \bigcup_{i_c \in \mathcal{I}_u^{cand}} \mathcal{D}_u^{\text{filtered}}(i_c)$

        \State \textbf{Augmentation:}
        \State $P(u) = \operatorname{Format}(\mathcal{I}_u^{cand}, \mathcal{D}_u^{\text{filtered}}, C)$ \Comment{Construct prompt}
        \State \textbf{Generation:}
        \State $\hat{y}_u = f_{\text{LLM}}(P(u); \theta_{\text{LLM}})$ \Comment{Generate recommendations using LLM}
        \State \textbf{Return:} $\hat{y}_u$
    \end{algorithmic}
\end{algorithm}
\vspace{-4pt}
The Algorithm~\ref{alg:cragru} describes CRAGRU's recommendation unlearning process, which primarily consists of three stages: retrieval, augmentation, and generation. In the retrieval stage, the algorithm first retrieves user-related interaction data, and then applies three different filtering strategies (user preference-based filtering, diversity and coverage-based filtering, and model-based filtering) to select key information to achieve unlearning. The augmentation stage integrates the filtered data, candidate items, and other auxiliary information into a prompt. Finally, the generation stage uses the prompt to call the LLM to generate recommendation results. The core of this algorithm is to effectively remove data that must be unlearned through filtering strategies, ensuring that the recommendations generated by the LLM meet the requirements of unlearning.

\end{document}